\documentclass[aps,prl,twocolumn,showpacs,preprintnumbers,nofootinbib,float,longbibliography,superscriptaddress]{revtex4-1}
\usepackage{epsfig}
\usepackage[dvipsnames]{xcolor}
\usepackage{float,amsmath,amssymb}
\usepackage{graphicx}
\usepackage{url}
\usepackage{bbold}
\usepackage[utf8]{inputenc}
\usepackage{physics}

\newcommand{\xt}{{\mathbf{x}_\perp}}

\usepackage[breaklinks,colorlinks,citecolor=citcolor,urlcolor=blue,linkcolor=lcolor]{hyperref}
\definecolor{lcolor}{rgb}{0.5,0,0}
\definecolor{citcolor}{rgb}{0,0.3,0.0}

\begin{document}

\title{Collision energy dependence in heavy ion collisions from nonlinear QCD evolution}

\author{Heikki M\"antysaari}
\affiliation{Department of Physics, University of Jyv\"askyl\"a, P.O. Box 35, 40014 University of Jyv\"askyl\"a, Finland}
\affiliation{Helsinki Institute of Physics, P.O. Box 64, 00014 University of Helsinki, Finland}

\author{Bj\"orn Schenke}
\affiliation{Physics Department, Brookhaven National Laboratory, Upton, NY 11973, USA}

\author{Chun Shen}
\affiliation{Department of Physics and Astronomy, Wayne State University, Detroit, Michigan 48201, USA}

\author{Wenbin Zhao}
\affiliation{Physics Department, University of California, Berkeley, California 94720, USA}
\affiliation{Nuclear Science Division, Lawrence Berkeley National Laboratory, Berkeley, California 94720, USA}

\begin{abstract}
We explore the effects of including the energy dependence determined from evolution equations within the color glass condensate framework on observables in ultra-relativistic heavy-ion collisions. This amounts to integrating the JIMWLK evolution equations into the IP-Glasma model, which is then coupled to viscous relativistic hydrodynamics. This methodology allows for a systematic representation of nuclei at specific Bjorken-$x$ values, which are probed at different center-of-mass energies of the collision and rapidities of final state particles. Comparing to the conventional IP-Glasma model, we find significant effects on multiplicity distributions and particle spectra, especially in smaller collision systems at the highest center of mass energies. Our results highlight the importance of incorporating nonlinear QCD evolution in the description of heavy ion collisions at varying center of mass energies, as the precise extraction of transport coefficients will be affected. This work establishes a robust framework for understanding the quark gluon plasma and nuclear structure at high energy, integrating small-$x$ physics into the initial conditions of heavy-ion collisions.
\end{abstract}

\maketitle

\noindent{\it 1. Introduction.}
Ultra-relativistic heavy-ion collisions probe nuclear matter under extreme conditions of temperature and density. At a high enough center-of-mass energy, $\sqrt{s_\mathrm{NN}}$, a transition to the quark-gluon plasma (QGP) occurs \cite{Muller:2012zq, Harris:2024aov}, and the produced matter behaves like a nearly perfect fluid \cite{Heinz:2013th,Gale:2013da,Jeon:2015dfa,Romatschke:2017ejr,Noronha:2024dtq}.
Relativistic viscous hydrodynamic models have been very successful in describing a wide range of bulk observables from both the Relativistic Heavy Ion Collider (RHIC) and the Large Hadron Collider (LHC), whose maximal center-of-mass energies differ by approximately a factor of 25; see e.g.~Ref.~\cite{Heinz:2024jwu} for a recent review. 

This energy dependence is usually incorporated in the initial state description. For example, in a Monte-Carlo Glauber model, one would adjust by hand the nucleon-nucleon cross section, relative contributions between wounded nucleons and binary collisions, as well as an overall normalization constant, to describe the energy-dependent particle production~\cite{Miller:2007ri}.
In the Eskola-Kajantie-Ruuskanen-Tuominen (EKRT) model~\cite{Kuha:2024kmq} the energy dependence arises via nuclear parton distribution functions fitted to deep inelastic scattering data.
The impact parameter-dependent Glasma (IP-Glasma) model~\cite{Schenke:2012wb,Schenke:2012hg}, based on the color glass condensate (CGC)~\cite{Iancu:2003xm} effective theory of quantum chromodynamics (QCD) at high energy, predicts the growth of particle production with energy based on the energy-dependent saturation scale $Q_s$, which is constrained within the impact parameter dependent saturation model (IPSat)~\cite{Bartels:2002cj,Kowalski:2003hm}.

These methods of describing the energy dependence of particle production in heavy-ion collisions have been rather successful, as the above mentioned models have been able to accurately reproduce many observables at various collision energies~\cite{Schenke:2020mbo,Heinz:2024jwu,Kuha:2024kmq}.
But as experimental observables become more complex and the precision of measurements improves, it is desirable to implement a more sophisticated description of the $\sqrt{s_\mathrm{NN}}$-dependence in the initial condition for hydrodynamic simulations.

In this work, we incorporate the energy evolution based on the perturbative JIMWLK equations~\cite{Jalilian-Marian:1996mkd,Jalilian-Marian:1997qno, Jalilian-Marian:1997jhx,Iancu:2001md, Ferreiro:2001qy, Iancu:2001ad, Iancu:2000hn,Mueller:2001uk}
into the IP-Glasma initial state description. This allows us to describe the $x$-dependence of the high-energy structure of the colliding nuclei (here $x$ is the fraction of the nucleon momentum carried by the parton).
The strategy is to initialize the description of incoming nuclei at a maximal $x=0.01$ (close to the value probed in RHIC kinematics) where the CGC approach is still applicable, and predict the nuclear gluon fields   at smaller $x$ by solving the JIMWLK equations. 
This way, we obtain the initial states for Au-Au collisions at RHIC, and Pb-Pb collisions at LHC, as well as O-O collisions at both colliders. 
In addition to producing an energy dependent saturation scale (a proxy for the nuclear density), this method includes the expected evolution of the nuclear geometry as a function of $x$, whose effects on multiplicity distributions and particle spectra are studied in this work. 

JIMWLK evolution has been shown to be applicable for computing inclusive quantities \cite{Gelis:2008rw} and included in heavy-ion initial state descriptions in the past to study the rapidity dependence of the initial geometry \cite{Schenke:2016ksl,Schenke:2022mjv,McDonald:2023qwc,Mantysaari:2023qsq,Mantysaari:2023prg,Mantysaari:2024qmt}. Here, we concentrate on bulk observables at midrapidity, and provide for the first time a study of different collision systems and energies using the perturbatively calculated initial state energy evolution. We argue that this energy evolution will affect the extraction of the QGP properties from global analyses.

\bigskip
\noindent {\it 2. Methodology.}
We use the IP-Glasma model to generate the initial color fields at $x=0.01$. This model is based on CGC effective theory and utilizes the classical description of gluon production first introduced in~\cite{Kovner:1995ja,Kovchegov:1997ke}, with numerical methods pioneered in~\cite{Krasnitz:1998ns,Krasnitz:1999wc,Krasnitz:2000gz,Lappi:2003bi}. The dependence of the saturation scale on the nuclear density and $x$ is given by the IPSat model, with parameters constrained by a fit to small-$x$ proton structure function data~\cite{Rezaeian:2012ji}.

The Wilson lines $V(\xt)$, which encode all information on the nuclear high-energy structure, are computed by solving the Yang-Mills equations for the gluon fields. One obtains
\begin{equation}
  V(\xt) = \mathrm{P}_{-}\left\{ \exp\left({-ig\int_{-\infty}^\infty \dd{z^{-}} \frac{\rho^a(z^-,\xt) t^a}{\boldsymbol{\nabla}^2 - m^2} }\right) \right\}\,,
  \label{eq:wline_regulated}
\end{equation}
where $\mathrm{P}_{-}$ represents path ordering in the $z^-$ direction, $\rho^a$ is the color charge density, and $t^a$ are the generators in the fundamental representation of SU(3), with $a$ being the color index. Here, we introduce an infrared regulator $m$, which is needed to avoid the emergence of unphysical Coulomb tails~\cite{Schenke:2012wb,Schenke:2012hg}. 

The Bjorken-$x$ dependence of the Wilson lines is obtained by numerically solving the JIMWLK evolution equation using methods described in Refs.~\cite{Mueller:2001uk,Lappi:2012vw,Rummukainen:2003ns,Cali:2021tsh}. We use a JIMWLK kernel with infrared regulation~\cite{Schlichting:2014ipa,Mantysaari:2022sux}, which is needed for finite size targets to avoid violation of the Froissart bound.
The gluon fields after the collision are then determined in terms of the JIMWLK-evolved Wilson lines, and the early-time evolution in the Glasma phase up to $\tau_0=0.4$ fm/c is obtained by solving the source-free Yang-Mills equations as implemented in IP-Glasma.

The free parameters in this initial state model determine the proton color charge density, its event-by-event fluctuating geometry~\cite{Mantysaari:2020axf}, and the JIMWLK evolution speed. Their values are taken from Ref.~\cite{Mantysaari:2022sux}. 
They were fitted to the available $\gamma+p \to \mathrm{J}/\psi+p$  data  from HERA~\cite{H1:2005dtp,H1:2013okq,ZEUS:2002wfj}  and from ultra-peripheral collisions~\cite{Bertulani:2005ru,Klein:2019qfb} measured at the LHC~\cite{ALICE:2014eof,ALICE:2018oyo,LHCb:2014acg, LHCb:2018rcm}, covering a wide range in $x$ from  $x\sim0.01$ down to $x\sim 10^{-6}$. 

Public codes that implement the IP-Glasma model and solve the JIMWLK evolution are available~\cite{ipglasma_code,jimwlk_code}. A version of the IP-Glasma that incorporates the JIMWLK evolution can be found from~\cite{ipglasma_jimwlk_code}.

In this work, we use two setups to describe the energy (Bjorken-$x$) dependence of the incoming nuclei.
Our main setup, denoted by ``{\texttt{+JIMWLK}}'', 
uses the Wilson lines evolved from the initial $x=0.01$ to $x=\left< p_T\right>e^{\pm y}/\sqrt{s_\mathrm{NN}}$ by solving the JIMWLK equation event-by-event (the different signs correspond to the different nuclei), with the $\left<p_T\right>$ being the charged hadron mean transverse momentum. We take $\left<p_T\right>=0.5$ GeV at RHIC kinematics and $\left<p_T\right>=0.65$ GeV at the LHC. 
In this work, we consider midrapidity kinematics $y=0$ for Pb-Pb, Au-Au, p-Au, and O-O collisions. 
For p-Pb collisions at 5.02 TeV, the midrapidity kinematics in the laboratory frame corresponds to $y=0.465$, leading to more evolution for the proton than for the Pb nucleus.
For comparison,  we use the ``{\texttt{$Q_s(x)$}}'' setup as a baseline, in which case the saturation scale depends on $x$ according to the IPsat parametrization as in the conventional IP-Glasma implementation, but other parameters of the IP-Glasma initial condition are taken at $x=0.01$.\footnote{Here, we follow the original implementation of IP-Glasma, with $x$ determined as described. We do not determine $Q_s$ iteratively by solving $x = x(\xt) = Q_s(x,\xt)/\sqrt{s_\mathrm{NN}}$, which has been done e.g.~in \cite{Schenke:2020mbo}. } By construction the two setups agree exactly at $x=0.01$, such that differences are expected to be small at RHIC energy.

The energy-momentum tensor $T^{\mu\nu}$ of the system is computed from the gluon fields at $\tau=\tau_0$, providing the initial condition for relativistic viscous hydrodynamic simulations performed using MUSIC~\cite{Schenke:2010nt,Schenke:2010rr,Paquet:2015lta}.

Because the strong coupling constant scales out of the Yang-Mills evolution (see discussion in \cite{Schenke:2020mbo}), we determine an overall scale factor $K$ for the system's energy-momentum tensor $T^{\mu\nu}$ by matching the charged hadron multiplicity in the 0--5\% most central Au-Au collisions at $\sqrt{s_\mathrm{NN}} = 200$ GeV. The JIMWLK evolution then predicts the collision energy dependence of particle production, which is distinct from the traditional approach with the Glauber model, where the normalization needs to be adjusted by hand for each center of mass energy.

As we are interested in midrapidity kinematics only, we use a boost invariant hydrodynamical evolution.
We employ a lattice-QCD-based equation of state~\cite{Bazavov:2014pvz, Moreland:2015dvc}, and consider both shear and bulk viscous effects during the hydrodynamic evolution by solving the Denicol-Niemi-Molnar-Rischke (DNMR) theory with spatial gradient terms up to the second order~\cite{Denicol:2012cn}.
The fluid cells are converted into hadrons on a constant energy density hypersurface with $e_\mathrm{sw} = 0.18$~GeV/fm$^3$ using the Cooper-Frye particlization prescription with Grad's 14-moment viscous corrections~\cite{Huovinen:2012is, Shen:2014vra, Zhao:2022ugy}. These hadrons are then fed into the hadronic transport model (UrQMD) for further scatterings and decay in the dilute hadronic phase~\cite{Bass:1998ca,Bleicher:1999xi}. For more details on the implementation, see Refs.~\cite{Schenke:2020mbo,Mantysaari:2024uwn} and references therein.

The hydrodynamical evolution of the QGP is sensitive to the matter properties, in particular to the (temperature dependent) shear and bulk viscosities. We use the following parametrization, and fix the free parameters by fitting to the measured charged hadron multiplicities \cite{PHENIX:2003iij}, mean transverse momentum \cite{STAR:2008med} and azimuthal anisotropy coefficients \cite{STAR:2015mki} in Au+Au collisions at $\sqrt{s_\mathrm{NN}}=200$ GeV:
\begin{align}
    \frac{\eta}{s}(T) &= \eta_0 + b(T - T_0) \Theta(T_0 - T) \\
    \frac{\zeta}{s}(T) &= \frac{(\zeta/s )_{\max}\Lambda^2}{\Lambda^2+ \left( T-T_0\right)^2},
    \Lambda = w_{\zeta} \left[1 + \lambda_{\zeta} {\rm sign} \left(T{-}T_0\right) \right]
\end{align}
The parameters used in this work are $\eta_0 = 0.14$, $T_0 = 0.18$ GeV, $b = -4$ GeV$^{-1}$, $(\zeta/s )_{\max} = 0.123$, $\lambda_{\zeta}$=-0.12, and $ w_{\zeta}=$ 0.05 GeV.

\begin{figure}
      \includegraphics[width=1.0\columnwidth]{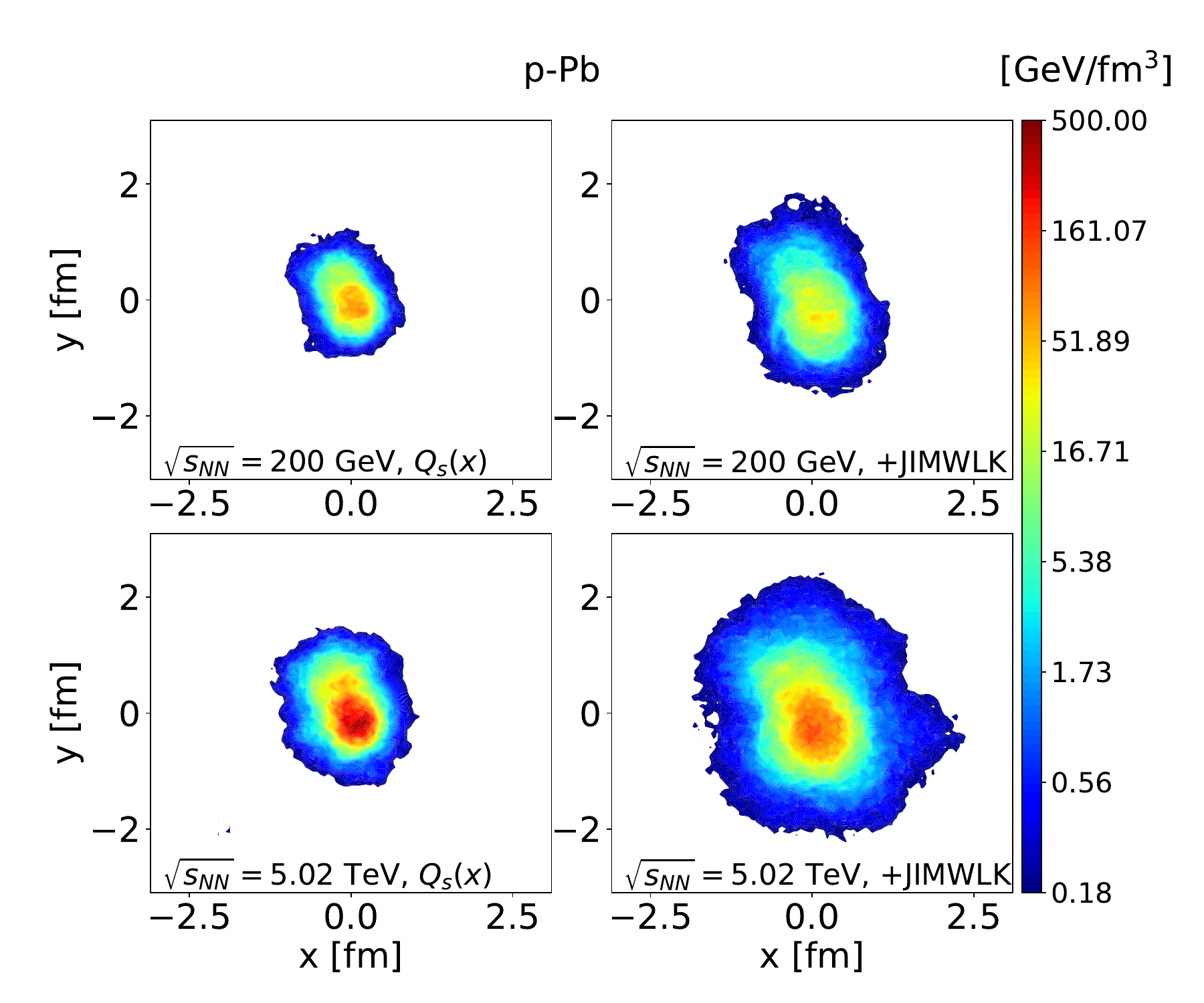}
    \caption{The initial energy density profiles at RHIC (top) and LHC (bottom) energies for p-Pb collisions obtained from ``\texttt{$Q_{s}(x)$}" (left) and ``\texttt{+JIMWLK}" (right) setups. Only contours above the freeze-out energy density of 0.18\,GeV/fm$^3$ are shown.} 
    \label{fig:ed}
\end{figure}
\begin{figure}
    \centering
    \includegraphics[width=\columnwidth]{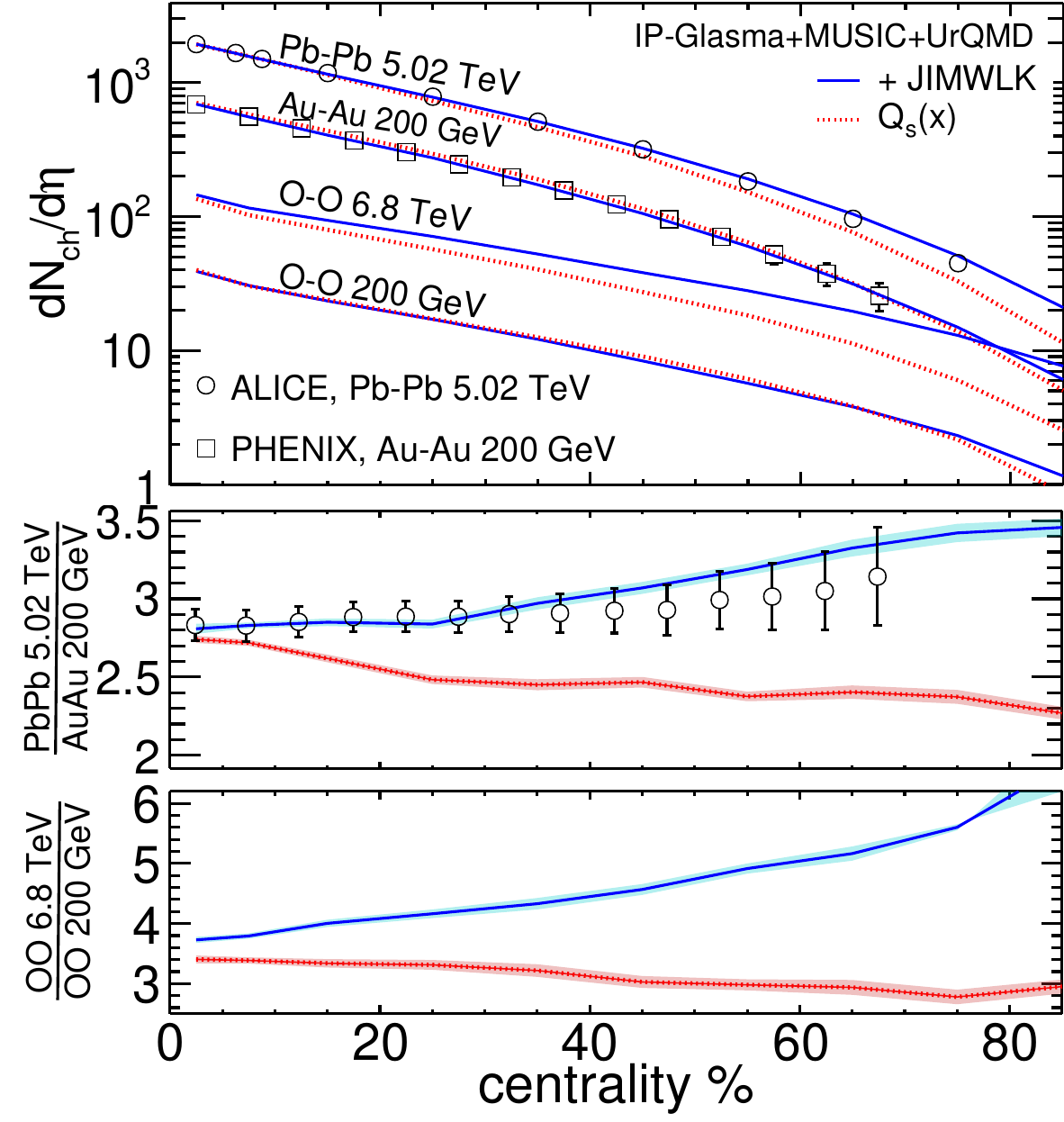}
    \caption{Charged hadron multiplicity distributions in Pb-Pb collisions at LHC and Au-Au collisions at RHIC energies, and O-O collisions at $\sqrt{s_\mathrm{NN}}=$ 6.8 TeV and 200 GeV. The experimental data is taken from \cite{ALICE:2015juo,PHENIX:2004vdg}.
     }
    \label{fig:dNchdeta}
\end{figure}

\bigskip
\noindent {\it 3. Signatures of high-energy evolution in the initial state.}
Figure~\ref{fig:ed} shows the initial energy density profiles in p-Pb collisions at $\tau_0=0.4$ fm/c for RHIC and LHC energies. We compare the results from JIMWLK evolution with those obtained by simply adjusting $Q_s$ according to the center of mass energy. We label the latter case as ``$Q_s(x)$''. It can be clearly observed that the JIMWLK energy evolution of the initial state increases its size. 
For example, with the complete JIMWLK energy evolution, the  transverse area of the initial energy density profile inside the freeze out surface  at midrapidity at the LHC collision energy is approximately 2.7 times that  without the JIMWLK energy evolution. 
This increase is caused by the effective growth of the proton size with energy evolution; for more discussion, see~\cite{Schlichting:2014ipa,Mantysaari:2018zdd,Mantysaari:2022sux}. Below we will show that such growth is evident in the multiplicity distribution and  mean transverse momentum observable in heavy-ion collisions.

Figure \ref{fig:dNchdeta} shows the centrality dependence of the charged hadron multiplicity in Pb-Pb collisions at 5.02 TeV, Au-Au collisions at 200 GeV, and O-O collisions at 6.8 TeV and 200 GeV. 
Here, the JIMWLK case has the same normalization factor $K=0.115$ at both RHIC and LHC energies, indicating that the JIMWLK energy evolution quantitatively captures the increasing energy density from RHIC to LHC. 
For the ``$Q_s(x)$'' case, we need to include a weak energy dependence to this factor and use $K_{\rm 200 GeV} = 0.106$ at RHIC and $K_{\rm 5020 GeV} = K_{\rm 6800 GeV} = 0.095$ at the LHC, suggesting that the $x$-dependence in the saturation scale alone cannot capture the full energy dependence.\footnote{Note that when determining $x$ and $Q_s(x)$ iteratively, as done e.g.~in \cite{Schenke:2020mbo}, the normalization does not need adjustment when changing collision energies.} 

The multiplicity distributions are almost identical for the two cases at RHIC energy. This is expected, as there is only limited evolution from the initial $x=0.01$ to $x=2.5\times 10^{-3}$ probed at $\sqrt{s_\mathrm{NN}}=200\,\mathrm{GeV}$.
At the LHC, one is probing the colliding nuclei at a much smaller $x=1.29 \times 10^{-4}$ (Pb-Pb) or $x = 7.35\times 10^{-5}$ (O-O), and a significant difference in the centrality dependence can be observed between the results with and without JIMWLK evolution.
In particular, the JIMWLK evolution results in multiplicity distributions decreasing more slowly as a function of centrality compared to the case where only the nuclear density (saturation scale) depends on the center-of-mass energy. 

The ALICE data prefers the JIMWLK case, where the JIMWLK evolution blurs the edges of the initial energy density, resulting in more overlapping areas and increased particle production in peripheral collisions.
We have checked that the observed changes in centrality dependence are not caused by differences in the viscous entropy production during the late-time hydrodynamic phase, and as such are initial state effects.\footnote{We note that in previous calculations using IP-Glasma+MUSIC+UrQMD \cite{Schenke:2020mbo}, which used the iterative determination of $Q_s(x)$, the centrality dependence of charged hadron production at LHC was also steeper than the experimental data.}

Predictions for O-O collisions are obtained by describing the  oxygen structure at $x=0.01$ using the variational Monte-Carlo method (VMC) from Ref.~\cite{Carlson:1997qn}. The O-O collisions provide a unique opportunity to directly examine the effect of the JIMWLK evolution of the initial state, as data from this collision system will be available from both RHIC and the LHC. This greatly reduces the uncertainties associated with the different number of participants in the same centrality class. Further, the system size is still relatively small, enhancing the effects of the geometry evolution, compared to Au-Au or Pb-Pb collisions.

Similar to Au-Au collisions, the energy evolution does not have sizable effects on the multiplicity distribution in O-O collisions at $\sqrt{s_\mathrm{NN}}=200$ GeV, providing a solid baseline for comparisons with the high-energy results. At $\sqrt{s_\mathrm{NN}}=6.8$ TeV, it can be clearly observed that the multiplicity distributions are highly sensitive to the energy evolution of the initial state. Specifically, $\dd N_\mathrm{ch}/\dd \eta$ vs.~centrality is again significantly flatter when the JIMWLK evolution is included.

The lower two panels in Fig.~\ref{fig:dNchdeta} show the ratios between the LHC Pb-Pb and RHIC Au-Au results, and between the reesults for O-O collisions at two energies. A clear enhancement in the multiplicity ratio is observed when the JIMWLK evolution is taken into account, compared to the approximate ``$Q_s(x)$'' setup. For example, in the 60-70\% centrality bin the ratio between Pb-Pb and Au-Au collisions with JIMWLK evolution is around 25\% larger than in the ``$Q_s(x)$'' setup. Including the energy evolution also shows a slightly increasing trend as a function of centrality, which is consistent with the experimental data and opposite to the trend observed when the evolution is approximated by an $x$-dependent saturation scale.

For the O-O collisions, the ratio between the 6.8 TeV and 200 GeV results shown in the bottom panel highlights the difference further. 
When the energy dependence is obtained by solving the JIMWLK equations, the ratio increases with increasing centrality, unlike in the case where only the saturation scale depends on $x$. This is exactly the same systematics as observed above in the case of Pb-Pb and  Au-Au collisions.
For example, in the 60--70\% centrality bin the ratio with JIMWLK evolution is approximately 70\% larger than in the $Q_s(x)$ case. 

As these are very significant effects, comparison to experimental data from RHIC and future LHC runs will be able to distinguish between the models, and potentially provide direct evidence for QCD evolution effects in heavy-ion collisions.

\begin{figure}
    \centering
    \includegraphics[width=\columnwidth]{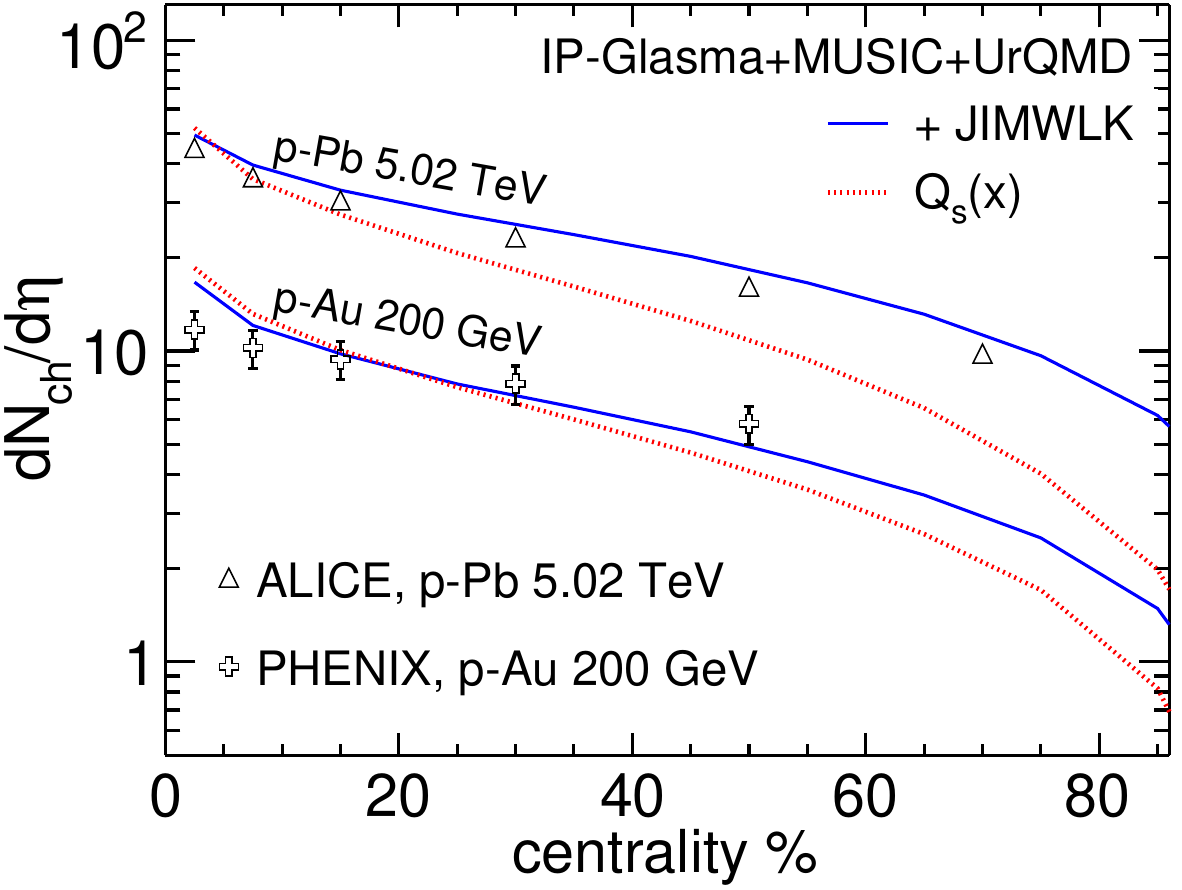}
    \caption{Charged hadron multiplicity distributions in p-Pb collisions at $\sqrt{s_\mathrm{NN}} = 5.02$ TeV and p-Au collisions at $\sqrt{s_\mathrm{NN}} = 200$ GeV. The experimental data is taken from \cite{ALICE:2013wgn,PHENIX:2003iij}.} 
    \label{fig:dNchdetapA}
\end{figure}

\begin{figure}
    \centering
    \includegraphics[width=\columnwidth]{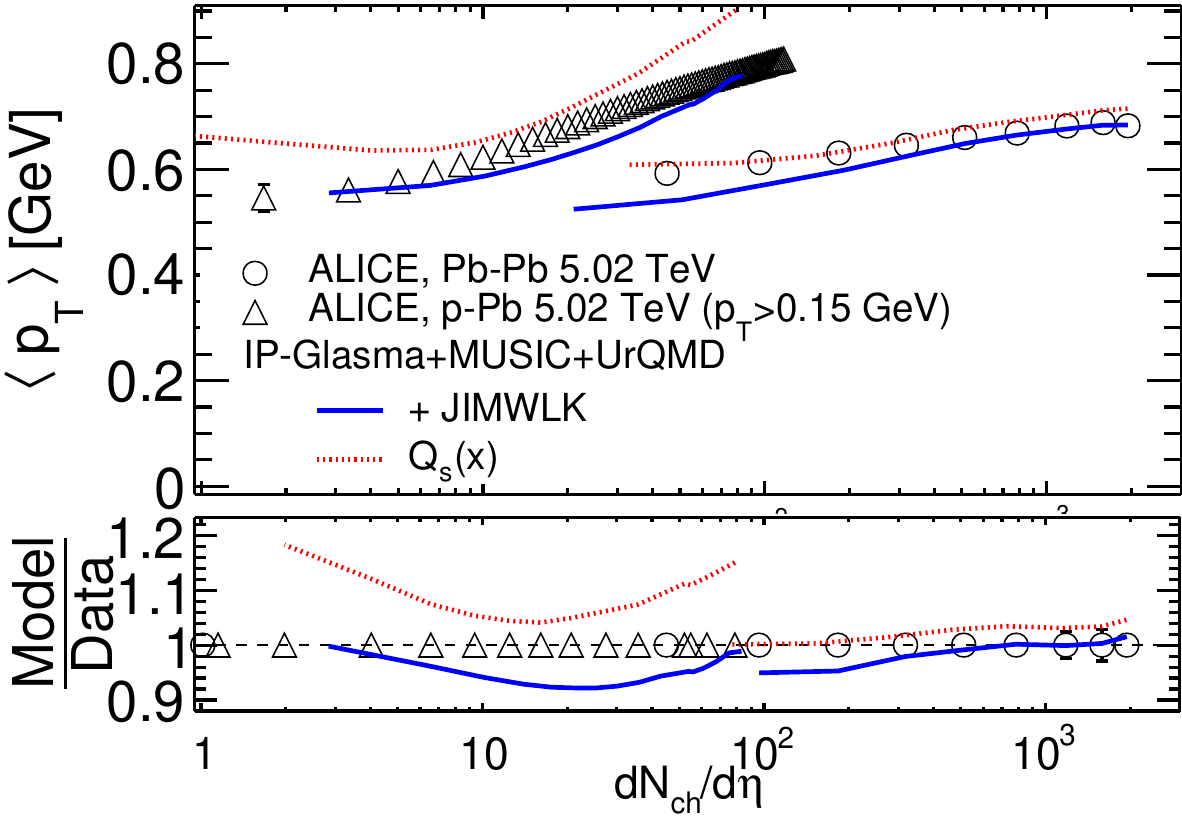}
    \caption{The charged hadron mean transverse momentum as a function of $\dd N_\mathrm{ch}/\dd \eta$ in Pb-Pb and p-Pb collisions at the LHC. The experimental data is taken from \cite{ALICE:2018hza}.}
    \label{fig:meanpt}
\end{figure}

Even larger differences can be observed in proton-nucleus (p-A) collisions. Figure \ref{fig:dNchdetapA} illustrates that, even at the RHIC energy of 200 GeV, a distinction between the two scenarios becomes evident. The smaller system is more sensitive to the energy-dependent geometry than the larger system, resulting in an improved description of the multiplicity distribution when the JIMWLK equation is used to describe the evolution from $x=0.01$ to $x=2.5\times 10^{-3}$.
As expected, these differences are amplified at the higher energies of the LHC. The JIMWLK evolution results in smoother and larger projectiles (see e.g.~\cite{Schlichting:2014ipa,Mantysaari:2023xcu}), which makes the multiplicity distribution flatter, leading to a better agreement with the experimental data from p-Pb collisions at the LHC.

The energy evolution of the initial gluon distribution also significantly impacts other observables. Figure \ref{fig:meanpt} shows the mean transverse momentum $\langle p_T \rangle$ in p-Pb and Pb-Pb collisions at $\sqrt{s_\mathrm{NN}}=5.02$ TeV. It is evident that with the JIMWLK evolution, $\langle p_T\rangle$ is notably reduced at LHC energies compared to the setup with an $x$-dependent saturation scale. This effect is particularly pronounced in p-Pb collisions and peripheral Pb-Pb collisions. The lower panel shows the ratio between the model predictions and experimental data, indicating that the inclusion of the JIMWLK evolution reduces $\langle p_T\rangle$ by approximately 20\% in p-Pb collisions.

This reduction can again be attributed to the growth of the proton (ion) size due to energy evolution, as depicted in Figure \ref{fig:ed}, which results in a broader initial energy density profile at small $x$. This leads to reduced pressure gradients, leading in turn to less radial flow buildup, finally lowering the mean $p_T$ of the final state hadrons.  As this change would affect extracted values of bulk viscosity when fitting to data, our findings underscore the importance of accounting for energy evolution when quantitatively extracting the transport coefficients of the QGP \cite{Bernhard:2016tnd,McDonald:2016vlt,Zhao:2017yhj,Auvinen:2017fjw,Nijs:2020ors,JETSCAPE:2020shq,Shen:2023awv,Virta:2024avu}.
We have also explored effects on anisotropic flow coefficients, yet we have found them to be smaller than those on the observables discussed above. We will report on a broader set of observables in an upcoming publication.

\bigskip
\noindent {\it 4. Summary. }
We investigated the effects of including nonlinear QCD evolution to describe the center of mass energy dependence of ultra-relativistic heavy-ion collisions. To do so, we incorporated JIMWLK evolution into the IP-Glasma initial state model, and performed viscous hydrodynamic calculations. All model parameters describing the initial state are constrained by $\gamma+p$ data. 

Comparing to calculations where the energy dependence is obtained only by adjusting the value of the saturation scale, our results demonstrate that the modifications of the spatial gluon distributions facilitated by JIMWLK evolution have significant effects on experimental 
observables. Specifically, the charged multiplicity distributions in Pb-Pb and p-Pb collisions at 5.02 TeV, and p-Au collisions at 200 GeV, are better described when including JIMWLK evolution, as it leads to the multiplicity distribution to decrease more slowly with centrality.

Predictions of the centrality dependence of $\dd N_{\rm ch}/\dd \eta$ in O-O collisions at $\sqrt{s_\mathrm{NN}}=200$ GeV and 6.8 TeV also show a strong sensitivity to the energy evolution of the initial state. The O-O collision system provides a unique opportunity as it will likely be the only system studied both at RHIC and LHC. This work makes testable QCD based predictions for how observables should evolve from RHIC to LHC energies.

The mean transverse momentum in p-Pb and Pb-Pb collisions at 5.02 TeV was found to be reduced when including JIMWLK evolution, particularly for smaller systems and peripheral heavy-ion collisions.

These findings underscore the importance of including the proper energy evolution in initial state models for accurately describing heavy-ion collisions. In the case of the color glass condensate based study presented here, the energy evolution broadens and smoothens the initial energy density profile, affecting particle production and final state flow observables. 

Our results establish a critical link between DIS measurements from HERA \cite{H1:2013okq,Mantysaari:2022ffw} and heavy ion data from RHIC and the LHC, facilitating a comprehensive future global Bayesian analysis to quantitatively extract transport coefficients of the QGP, nuclear structure parameters, and to identify signals of gluon saturation.  

\bigskip
\noindent {\it{Acknowledgments.}}
B.P.S. and C.S. are supported by the U.S. Department of Energy, Office of Science, Office of Nuclear Physics, under DOE Contract No.~DE-SC0012704 and Award No.~DE-SC0021969, respectively.  C.S. acknowledges a DOE Office of Science Early Career Award. This material is based upon work supported by the U.S. Department of Energy, Office of Science, Office of Nuclear Physics, within the framework of the Saturated Glue (SURGE) Topical Theory Collaboration.
H.M. is supported by the Research Council of Finland, the Centre of Excellence in Quark Matter, and projects 338263 and 359902, and under the European Research Council (ERC, grant agreements No. ERC-2023-101123801 GlueSatLight and No. ERC-2018-ADG-835105 YoctoLHC).
W.B.Z. is supported by the National Science Foundation (NSF) under grant number ACI-2004571 within the framework of the XSCAPE project of the JETSCAPE collaboration.
This research was done using resources provided by the Open Science Grid (OSG)~\cite{Pordes:2007zzb, Sfiligoi:2009cct}, which is supported by the National Science Foundation award \#2030508.
The content of this article does not reflect the official opinion of the European Union and responsibility for the information and views expressed therein lies entirely with the authors.
\bibliography{refs}

\end{document}